\documentclass[sigconf]{acmart}

\usepackage{booktabs}
\usepackage{cleveref}
\usepackage{dirtytalk}
\usepackage{natbib}
\usepackage{subfig}
\usepackage{enumitem}
\usepackage{url}
\usepackage{bbding} 
\usepackage{framed}

\begin{document}

\title{Studying Attitudes and Social Norms in Agile Software Development}

\author{Lucas Gren}
\orcid{1234-5678-9012}
\affiliation{%
  \institution{Chalmers University of Technology and The University of Gothenburg\\}
  \streetaddress{The Department of Computer Science and Engineering}
  \city{Gothenburg} 
  \state{Sweden} 
  \postcode{412--92}
}
\email{lucas.gren@cse.gu.se}

\begin{abstract}
The purpose of this paper is to review research on attitudes and social norms and connect it to the agile software development context. Furthermore, I propose additional theories from social psychology (mainly the theory of planned behavior and using the degree of internalization of social norms) that would most certainly be useful for further sense-making of human factors-related research on agile teams. 
\end{abstract}


\copyrightyear{2018} 
\acmYear{2018} 
\setcopyright{acmlicensed}
\acmConference[XP '18 Companion]{XP '18 Companion}{May 21--25, 2018}{Porto, Portugal}
\acmPrice{15.00}
\acmDOI{10.1145/3234152.3234157}
\acmISBN{978-1-4503-6422-5/18/05}

%
%

%
%



\maketitle

\section{Introduction}\label{sec:introduction}
Two years ago, \citet{stray2016exploring} published a first study finding and categorizing social norms in agile software development teams. Social norms are often divided into injunctive (prescriptive) and descriptive ones \citep{reno1993transsituational} --- meaning norms that indicate what most people \emph{are} doing, and what members of the team \emph{ought} to do, --- which has also gained empirical support \citep{schultz2007constructive}. \citet{teh} showed that productive social norms lead to increased performance in a more general small study on software development teams, which showed that they are of utter importance for high quality development processes, especially in the agile team-based context. The authors \citet{stray2016exploring} suggest the use of field observations in order to understand social norms in agile software teams better. In addition, they suggest an investigation of social values in order to reveal underlying assumptions to certain behavior. I believe the research agenda on understanding behavior in agile software development teams will have increased usefulness if attitudes are studied in connection to social norms. Also, research on agile software teams should be supported by some more definitions and theories of attitudes and norms from social psychology. Suggesting such theories and approaches to research in the agile software development context is the goal of this paper.

\section{Attitudes}\label{sec:att}
\subsection{The History of Research on Attitudes}
According to \citet{hogg2014sp}, attitude research has gone through many different stages for almost a hundred years. The simplest one-component model of attitudes is to see them as an affect (positive or negative) towards, or an evaluation of, an object \citep{thurstone1931measurement}. However, such a model had issues with predicting behavior \citep{wicker1969attitudes}, which is the most important and practical usage of even investigating an attitude toward an object. The two-component model added the mental readiness to act to the construct of attitudes, since such an addition showed to be essential for the probability to actually act on an attitude \citep{hogg2014sp}. Yet another addition is the three-component model, proposing that attitudes consists of affect, cognitive, and behavioral components \citep{eagly1978attitudes}. The definition of what attitudes are is a thorny issue, but includes the following three aspects: 1) Attitudes are relatively permanent in contrast to momentary feelings, 2) They are limited to socially significant events or objects, and 3) They are somewhat generalizable in that they are towards a category of events and objects and not exclusive ones at only one point in time. Therefore, an attitude is ``made up of thought and ideas, a cluster of feelings, likes and dislikes, and behavioral intentions'' \citep{hogg2014sp}, which makes them intimately connected to the psychological definition of ``schema'' that are cognitive structures of knowledge and attributes of psychological objects. The idea is that an attitude saves cognitive energy since it is based on previous knowledge (or, more correctly, beliefs). Even though it is difficult to obtain an exact definition of attitudes, they are of great value in that they do predict behavior if the mechanisms, triggers, and context are controlled for \citep{hogg2014sp}. 

In gaining more predictive power, modeling the interaction between attitudes, beliefs, and behavioral intentions can better explain why an actual behavior is triggered or not. For example, social norms may play a critical role in the attitude-behavior relations and completely ruin our predictions if not well understood \citep{ajzen2001nature}. In a meta-study by \citet{kraus1995attitudes} the overall effect size of predicting behavior was 14.4\%, which means that 14.4\% in the response variable (i.e.\ behavior) could be explained by the factor (i.e.\ measurement of attitudes). This number might be perceived as low, however, in such complex systems the number shows that attitudes are a highly relevant construct to use. In addition, the closer the attitudes are to specific behavior, the better predictors they are \citep{kraus1995attitudes}. 

\subsection{The Theory of Planned Behavior} \label{tpb}
\begin{figure}
\label{planned}
\includegraphics[scale=0.54]{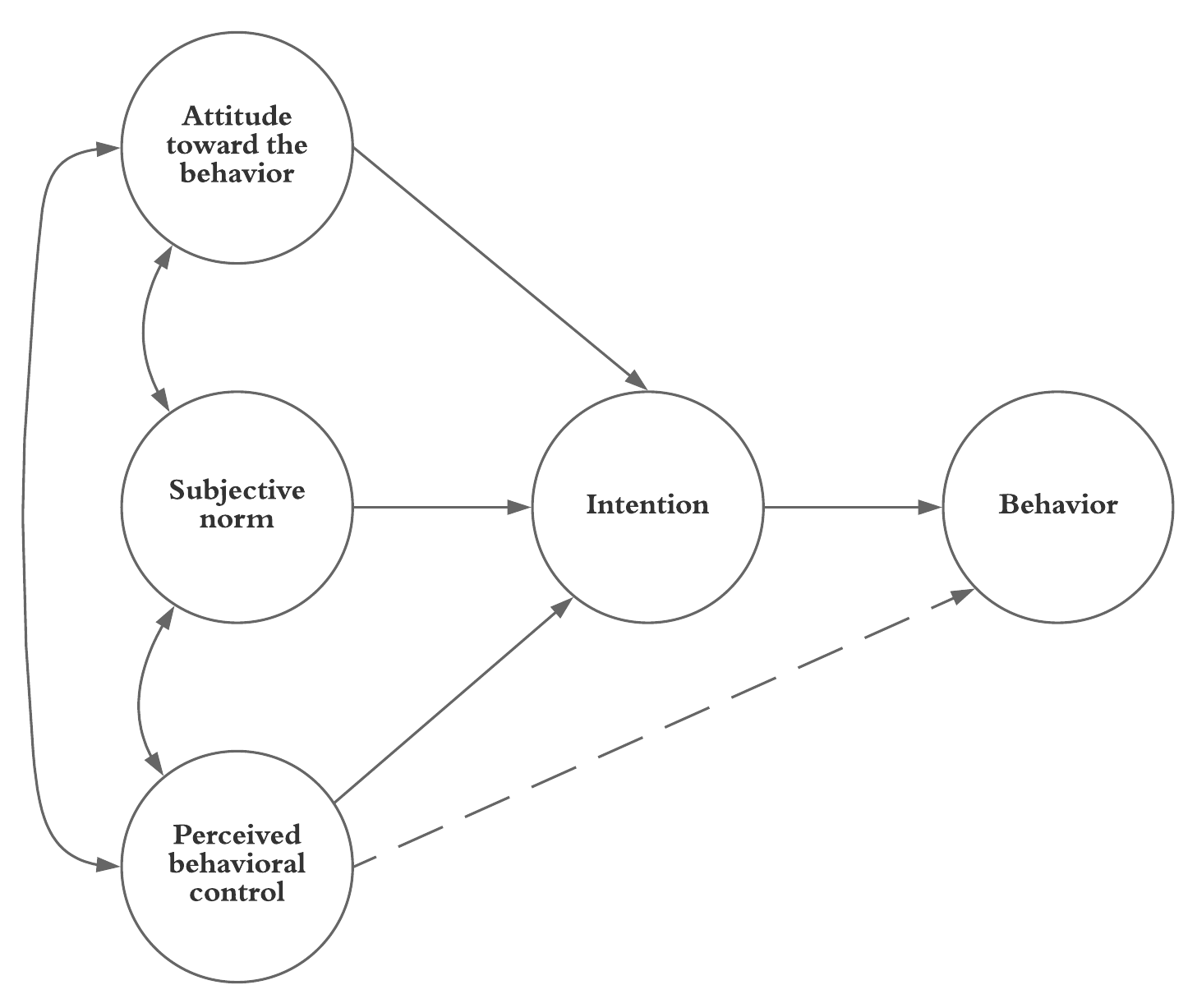}
\caption{Theory of Planned Behavior.}
\end{figure}

An integrated model of attitudes (and possibly the most famous) is the theory of planned behavior \citep{ajzen}. This theory comprises the following three parts: belief, intention, and action. The belief part can be divided into three subcategories: subjective norm, attitude towards behavior, and perceived behavioral control. Intention is determined by the three types of belief and consists of only the behavioral intention. The behavioral intention then leads to action, which is the actual behavior that can also be directly affected by the perceived behavioral control, according to \citet{ajzen}, see Figure~\ref{planned}. What distinguishes the theory of planned behavior from the predecessor, the theory of reasoned action, is the addition of the perceived behavioral control. For example, if the predicted behavior is ``sharing issues in the daily stand-up,'' the attitude towards sharing issues could be ``I think it is important to share issues in development projects,'' and the subjective norm could be an internalized norm in the social context that sharing issues in the daily stand-up is something of value which proves us as good individuals in relation to each other. We must also assess the intention ``to share issues in the daily stand-up,'' and not more general proxies like ``sharing issues'' or ``talking during daily stand-ups.'' We also need to ascertain perceived control over the possibility to ``share issues in the daily stand-up'' if we are to have good predictability of the behavior \citep{ajzen}.

\subsection{Attitude Functions}
At the core of the defined functionality of attitudes, attitudes are essential because they facilitate human adaptation to the environment \citep{ajzen2001nature}. \citet{ajzen2001nature} specifically lists the following as prominent functions in theory: 1) The value-expressive function, 2) The knowledge function, 3) The ego-defensive function, 4) The social-adjustive function, and 5) The utilitarian function. 

The value-expressive function is simply the fact that an attitude can serve as a means to express a human value. One example from the agile software development context could be positive attitudes toward the idea of sustainable pace. Advocating this practice both in expression and intended behavior, could be a way for developers to express their values of respecting a professionals personal life. 

Since attitudes are intimately connected to schema and are represented in memory \citep{olson1993attitudes}, their function as knowledge is natural. \citet{verplanken1999habit} argue that affective evaluations are faster than cognitive evaluations, which means that an affect-triggering attitude can retrieve knowledge faster. Attitudes can be seen as connections between an object and an evaluation of it, and therefore, strong object-evaluation associations are functional in that they help us access information and knowledge. One example from the agile software development context could be negative attitudes towards commercial software. If a developer is used to open-source, attitudes towards commercial software will retrieve knowledge about that in comparison with open-source. This will trigger fast retrieval of knowledge, which could be useful in negotiation or argumentation.

Attitudes that protect self-esteem are of utter importance since failing to filter the world in order to maintain self-esteem often lead to psychological illness and depression \citep{hogg2014sp}. An example from the agile software development context could be a negative attitude towards stating detailed and negative feedback to colleagues openly. This could then protect the developer against getting lower self-esteem if such feedback is given, since (s)he can attribute such comments to irrelevant political behavior, and therefore, does not have to change her\slash his own behavior based on such feedback.

The social-adjustive function simplifies our lives since we can match attitudes with other people and thereby find people we can relate to. However, a distinction is often made between people who are low self-monitoring (mostly have attitudes as value-expressions) and whose who have high self-monitoring (mostly have attitudes for social adjustment) \citep{debono1987investigating}. However, even for a developer who is more of the former, a useful way of navigating through a social context with people, could be to have attitudes for social-adjustive purposes. An example could be an attitude that people a developer relates a lot to always pairing up with new team-members. If such behavior is seen, or is not seen, a person could then adjust the social behavior and also make sure to do the same in that very context. 

Some researchers claim that the main functionality of attitudes is that of utility. Merely having an attitude (i.e.\ an accessible evaluation, or appraisal, of an object) is useful \citep{hogg2014sp}. According to a meta-study by \citet{glasman2006forming}, the following four factors have an effect on the strength of the correlation between the attitude and behavior: 1) The attitudes are accessible, 2) The attitudes are stable over time, 3) People have had direct experience with the attitude object, and 4) People frequently report their attitudes.

\subsection{Attitude Research in Software Engineering}
I have only found a couple of studies within software engineering that apply the construct of attitudes in some form. First, the study by \citet{passos2013beliefs} uses the theory of planned behavior to characterize a belief systems of a set of software development teams from three different companies. It was mainly used to explain the teams' behavior based on participants' beliefs and attitudes toward new software development practices, however with a very small sample of nine participants. The researchers concluded that the framework was useful for explaining the described behavior, but much more research is needed in the software engineering context. 

In a study by \citet{Lenberg2016} they investigated which antecedents affect attitudes towards organizational change in the software engineering context using industrial data from one company. They concluded that: ``knowledge about the intended change outcome, their understanding of the need for change, and their
feelings of participation in the change process'' \citep{Lenberg2016} are of importance. Furthermore, they concluded that the attitudes towards ``openness to change'' was predicted by the three
underlying concepts (knowledge, need for change, and participation), while the attitude concept ``readiness for change'' was predicted by the ``need for change'' and ``participation.'' The authors did not investigate social team norms in relation to the attitudes, but conclude that this aspect need understanding since group norms that are contradicting the attitudes will hamper even an intended behavior. 

In a study by \citet{feldt2010}, the authors investigated software engineers general work attitudes and their connection to results on a personality test. They concluded that general work attitudes were connected more to openness and extroversion in software engineering than other types of personalities. Later studies have shown that personality tests have very little predictive value (see e.g.\ \cite{forty}) and I believe looking at behavior in context is a much better idea if prediction is the goal. 

As have been shown, social norms are one of the key causes of behavior and need to be understood when investigating attitudes. Therefore, social norms are defined and explained next.

\section{Social Norms}\label{sec:norms}
\citet{hogg2014sp} define \emph{norms} as:

\begin{quote}
``Attitudinal and behavioral uniformities that define group membership and differentiate between groups.'' 
\end{quote}
\citet{schultz2007constructive} nicely show the difference between descriptive and injunctive norms, but by using a quite simple behavioral change (i.e.\ saving energy at home). They showed that households only receiving normative messages of their energy consumption would either spend more or less energy depending on where they were in relation to the normative consumption. This caused a ``boomerang effect'' since some households increased the consumption after the information was given to them. In order to also add an injunctive message another group received a sad or happy emoticon in connection with their energy consumption information that eliminated the boomerang effect. While their experiment showed that there are differences between descriptive and injunctive norms, just like \citet{hogg2014sp} also suggest, the dynamic nature of groups norms are often more complex and therefore the descriptive and prescriptive aspects of norms are often interrelated, as also found in the agile team context \citep{stray2016exploring}.

Norms are described as somewhat tacit since they are rules set on group-level that diverge from personal self-interest only. \citet{hogg2014sp} also highlight research supporting the fact that norms are tied to groups and not individuals, e.g.\ norms can influence individuals even in the physical absence of the group. With this group perspective of norms, the classic division of norms into descriptive and injunctive categories become more intertwined, since norms often are prescriptive in groups. More than anything, norms provide a frame of reference that reduce individual uncertainty and facilitate knowing how to behave, i.e.\ not only in reference to how the majority behaves \citep{hogg2014sp}.

\citet{thogersen2006norms} also believes in the need for higher resolution in making distinctions between descriptive and injunctive norms. He suggested dividing injunctive norms into more categories that he calls personal or subjective social norms. The personal norms are then further divided into introjected and integrated norms depending on their degree of internalization. The internalization is in relation to how much the individual is part of that group norm and not really a norm on individual level per se. It seems to be a fact that social psychology in general has moved from only believing that the individual matters \citep{hogg2014sp}, to an over-belief in the power of descriptive norms in modern environmental marketing \citep{schultz2007constructive} to again emphasizing the importance of the degree of internalization \cite{thogersen2006norms}. In relation to agile teams, I do think that some of these experiments are too simplistic in that they do not take the type of task into consideration. Influencing a purchasing decision when a consumer makes a decision alone in comparison to when an agile team gets to make the same decision. I would imagine the descriptive or subjective social norms would be much stronger with face-to-face peer pressure than all these environmentally friendly individual purchasing studies. 

I have now introduced the definition and function of social norms based on social psychological research findings. Understanding attitudes and norms in the software engineering context is a first step that surely needs more research focus, however, the practical usefulness lies mostly in how to change attitudes and norms. How to achieve such changes shapes the last parts of the discussion of this paper.


\section{Discussion}\label{sec:discussion}
In the main text of this paper, I have shown how attitudes and social norms work based on social psychological research findings. As values are more high-level beliefs about the social context \citep{schwartz1992universals}, they are trickier to measure and study their direct effects on behavior, however they can be measured in relation to what influences attitudes. Therefore, I believe focus should be on attitudes and the factors influencing them as presented in the theory of planned behavior \citep{ajzen}. Understanding what drives behavior in software engineering teams is of utter importance in order to optimize and improve the development process of any team. Instead of re-inventing the wheel, software engineering research should draw upon more general social psychology finding and use the software engineering context as a new application domain. The results from investigating new complex and innovative projects (like the agile team context) is surely of interest to general social psychology, since such finding provide a new context for such research. It is somewhat a pity that most research on behavior in the software engineering domain is conducted by software engineering researchers interested in psychology and not as well social or organizational psychologist interested in software engineering. Surely, both perspectives are needed in order to optimize the useful output of research resources spent in the field. 

To finalize this paper, the practical interest in changing attitudes and norms in the software engineering context could also draw upon attitude and norm change research already conducted in social psychology. 

One popular way of looking at attitude change is the dual-process models of persuasion. Humans seem to often select between thinking strategies depending on how much cognitive energy we have for the matter at hand. The energy is dependent on the elaboration likelihood, which relates to if we are motivated to spend cognitive energy on the message \citep{petty1986elaboration}. If the content of the message is important to us we thoroughly assess the content and if it is favorable to us, we change our attitude (or form one). In contrast, if the content is unimportant to us, we look for so-called ``peripheral cues.'' Such cues are based on superficial aspects such as mood, attractive or expert communicator, number of arguments, etc.\ \citep{petty1986elaboration}. There is also a family of theories that stress that people try to maintain internal consistency, order, and agreement among their various types of cognition. When we get new information we try to match it with what we already know and feel, and if there is a dissonance, we either reject the information presented to us, or change our attitude towards the object (such theories are called cognitive dissonance theories) \citep{hogg2014sp}. I would like to remind the reader here about the drivers behind the attitude-behavior relationship in Section~\ref{tpb}, where also an example was given from the agile team context.

In order to change social norms using a normative intervention, one should use different strategies depending on the context. In order to change social norms regarding agile leadership and management, one would have to focus more on the internalization of agile values at universities in order to affect software engineering students over a number of years to have these agile and more democratic values in their daily work. On the other hand, for people who do not have these values internalized, using descriptive normative change management would not be the best strategy, but instead maybe organize compulsory ``agile councils'' in the organization where everybody is obliged to participate in how to change the situation locally in each team (i.e.\ trying to maximize subjective social norm change). In order to simplify this rather complex psychological reasoning, it is better to show employees good agile team example and put pressure on people to change though showing the desired behavior, instead of communicating how most agile teams behave and expect the ones that do not, to change because of that information only. 

As a closing remark, some studies have proposed the use of social identity theory \citep{hogg2000management} to re-conceptualize the the role of norms in attitude-behavior relations, which could simplify the measurement of such influence \citep{terry1996group}.

\section{Conclusion}
In conclusion, I have provided some definitions and research findings from social psychology that I believe are valuable when investigating behavior in the agile software development context. Further research endeavors should draw upon this work in order to make sense of the human factors related to agile teams in the development process.

\bibliographystyle{ACM-Reference-Format}
\bibliography{refssocial}  


\begin{thebibliography}{00}


\ifx \showCODEN    \undefined \def \showCODEN     #1{\unskip}     \fi
\ifx \showDOI      \undefined \def \showDOI       #1{{\tt DOI:}\penalty0{#1}\ }
  \fi
\ifx \showISBNx    \undefined \def \showISBNx     #1{\unskip}     \fi
\ifx \showISBNxiii \undefined \def \showISBNxiii  #1{\unskip}     \fi
\ifx \showISSN     \undefined \def \showISSN      #1{\unskip}     \fi
\ifx \showLCCN     \undefined \def \showLCCN      #1{\unskip}     \fi
\ifx \shownote     \undefined \def \shownote      #1{#1}          \fi
\ifx \showarticletitle \undefined \def \showarticletitle #1{#1}   \fi
\ifx \showURL      \undefined \def \showURL       #1{#1}          \fi
\providecommand\bibfield[2]{#2}
\providecommand\bibinfo[2]{#2}
\providecommand\natexlab[1]{#1}
\providecommand\showeprint[2][]{arXiv:#2}

\bibitem[\protect\citeauthoryear{Ajzen}{Ajzen}{1991}]%
        {ajzen}
\bibfield{author}{\bibinfo{person}{Icek Ajzen}.}
  \bibinfo{year}{1991}\natexlab{}.
\newblock \showarticletitle{The theory of planned behavior}.
\newblock \bibinfo{journal}{{\em Organizational behavior and human decision
  processes\/}} \bibinfo{volume}{50}, \bibinfo{number}{2}
  (\bibinfo{year}{1991}), \bibinfo{pages}{179--211}.
\newblock


\bibitem[\protect\citeauthoryear{Ajzen}{Ajzen}{2001}]%
        {ajzen2001nature}
\bibfield{author}{\bibinfo{person}{Icek Ajzen}.}
  \bibinfo{year}{2001}\natexlab{}.
\newblock \showarticletitle{Nature and operation of attitudes}.
\newblock \bibinfo{journal}{{\em Annual review of psychology\/}}
  \bibinfo{volume}{52}, \bibinfo{number}{1} (\bibinfo{year}{2001}),
  \bibinfo{pages}{27--58}.
\newblock


\bibitem[\protect\citeauthoryear{Cruz, da~Silva, and Capretz}{Cruz
  et~al\mbox{.}}{2015}]%
        {forty}
\bibfield{author}{\bibinfo{person}{Shirley Cruz}, \bibinfo{person}{Fabio da
  Silva}, {and} \bibinfo{person}{Luiz Capretz}.}
  \bibinfo{year}{2015}\natexlab{}.
\newblock \showarticletitle{Forty years of research on personality in software
  engineering: {A} mapping study}.
\newblock \bibinfo{journal}{{\em Computers in Human Behavior\/}}
  \bibinfo{volume}{46} (\bibinfo{year}{2015}), \bibinfo{pages}{94--113}.
\newblock


\bibitem[\protect\citeauthoryear{DeBono}{DeBono}{1987}]%
        {debono1987investigating}
\bibfield{author}{\bibinfo{person}{Kenneth~G DeBono}.}
  \bibinfo{year}{1987}\natexlab{}.
\newblock \showarticletitle{Investigating the social-adjustive and
  value-expressive functions of attitudes: {I}mplications for persuasion
  processes}.
\newblock \bibinfo{journal}{{\em Journal of Personality and Social
  Psychology\/}} \bibinfo{volume}{52}, \bibinfo{number}{2}
  (\bibinfo{year}{1987}), \bibinfo{pages}{279}.
\newblock


\bibitem[\protect\citeauthoryear{Eagly and Himmelfarb}{Eagly and
  Himmelfarb}{1978}]%
        {eagly1978attitudes}
\bibfield{author}{\bibinfo{person}{Alice~H Eagly} {and} \bibinfo{person}{Samuel
  Himmelfarb}.} \bibinfo{year}{1978}\natexlab{}.
\newblock \showarticletitle{Attitudes and opinions}.
\newblock \bibinfo{journal}{{\em Annual review of psychology\/}}
  \bibinfo{volume}{29}, \bibinfo{number}{1} (\bibinfo{year}{1978}),
  \bibinfo{pages}{517--554}.
\newblock


\bibitem[\protect\citeauthoryear{Feldt, Angelis, Torkar, and Samuelsson}{Feldt
  et~al\mbox{.}}{2010}]%
        {feldt2010}
\bibfield{author}{\bibinfo{person}{Robert Feldt}, \bibinfo{person}{Lefteris
  Angelis}, \bibinfo{person}{Richard Torkar}, {and} \bibinfo{person}{Maria
  Samuelsson}.} \bibinfo{year}{2010}\natexlab{}.
\newblock \showarticletitle{Links between the personalities, views and
  attitudes of software engineers}.
\newblock \bibinfo{journal}{{\em Information and Software Technology\/}}
  \bibinfo{volume}{52}, \bibinfo{number}{6} (\bibinfo{year}{2010}),
  \bibinfo{pages}{611--624}.
\newblock


\bibitem[\protect\citeauthoryear{Glasman and Albarracin}{Glasman and
  Albarracin}{2006}]%
        {glasman2006forming}
\bibfield{author}{\bibinfo{person}{Laura~R Glasman} {and}
  \bibinfo{person}{Dolores Albarracin}.} \bibinfo{year}{2006}\natexlab{}.
\newblock \showarticletitle{Forming attitudes that predict future behavior: {A}
  meta-analysis of the attitude-behavior relation}.
\newblock \bibinfo{journal}{{\em Psychological bulletin\/}}
  \bibinfo{volume}{132}, \bibinfo{number}{5} (\bibinfo{year}{2006}),
  \bibinfo{pages}{778}.
\newblock


\bibitem[\protect\citeauthoryear{Hogg and Terry}{Hogg and Terry}{2000}]%
        {hogg2000management}
\bibfield{author}{\bibinfo{person}{Michael~A Hogg} {and}
  \bibinfo{person}{Deborah~I Terry}.} \bibinfo{year}{2000}\natexlab{}.
\newblock \showarticletitle{Social identity and self-categorization processes
  in organizational contexts}.
\newblock \bibinfo{journal}{{\em Academy of management review\/}}
  \bibinfo{volume}{25}, \bibinfo{number}{1} (\bibinfo{year}{2000}),
  \bibinfo{pages}{121--140}.
\newblock


\bibitem[\protect\citeauthoryear{Hogg and Vaughan}{Hogg and Vaughan}{2014}]%
        {hogg2014sp}
\bibfield{author}{\bibinfo{person}{Michael~A. Hogg} {and}
  \bibinfo{person}{Graham~M. Vaughan}.} \bibinfo{year}{2014}\natexlab{}.
\newblock \bibinfo{booktitle}{{\em Social Psychology\/} (\bibinfo{edition}{7}
  ed.)}.
\newblock \bibinfo{publisher}{Pearson}, \bibinfo{address}{Harlow, England}.
\newblock


\bibitem[\protect\citeauthoryear{Kraus}{Kraus}{1995}]%
        {kraus1995attitudes}
\bibfield{author}{\bibinfo{person}{Stephen~J Kraus}.}
  \bibinfo{year}{1995}\natexlab{}.
\newblock \showarticletitle{Attitudes and the prediction of behavior: {A}
  meta-analysis of the empirical literature}.
\newblock \bibinfo{journal}{{\em Personality and social psychology bulletin\/}}
  \bibinfo{volume}{21}, \bibinfo{number}{1} (\bibinfo{year}{1995}),
  \bibinfo{pages}{58--75}.
\newblock


\bibitem[\protect\citeauthoryear{Lenberg, Wallgren~Tengberg, and Feldt}{Lenberg
  et~al\mbox{.}}{2016}]%
        {Lenberg2016}
\bibfield{author}{\bibinfo{person}{Per Lenberg},
  \bibinfo{person}{Lars~G{\"o}ran Wallgren~Tengberg}, {and}
  \bibinfo{person}{Robert Feldt}.} \bibinfo{year}{2016}\natexlab{}.
\newblock \showarticletitle{An initial analysis of software engineers'
  attitudes towards organizational change}.
\newblock \bibinfo{journal}{{\em Empirical Software Engineering\/}}
  (\bibinfo{year}{2016}).
\newblock
\showDOI{%
\url{http://dx.doi.org/10.1007/s10664-016-9482-0}}


\bibitem[\protect\citeauthoryear{Olson and Zanna}{Olson and Zanna}{1993}]%
        {olson1993attitudes}
\bibfield{author}{\bibinfo{person}{James~M Olson} {and} \bibinfo{person}{Mark~P
  Zanna}.} \bibinfo{year}{1993}\natexlab{}.
\newblock \showarticletitle{Attitudes and attitude change}.
\newblock \bibinfo{journal}{{\em Annual review of psychology\/}}
  \bibinfo{volume}{44}, \bibinfo{number}{1} (\bibinfo{year}{1993}),
  \bibinfo{pages}{117--154}.
\newblock


\bibitem[\protect\citeauthoryear{Passos, Cruzes, and Mendon{\c{c}}a}{Passos
  et~al\mbox{.}}{2013}]%
        {passos2013beliefs}
\bibfield{author}{\bibinfo{person}{Carol Passos}, \bibinfo{person}{Daniela~S
  Cruzes}, {and} \bibinfo{person}{Manoel Mendon{\c{c}}a}.}
  \bibinfo{year}{2013}\natexlab{}.
\newblock \showarticletitle{Beliefs underlying teams intention and practice:
  {A}n application of the theory of planned behavior}.
\newblock \bibinfo{journal}{{\em Proceeding of ESELAW\/}}  \bibinfo{volume}{13}
  (\bibinfo{year}{2013}).
\newblock


\bibitem[\protect\citeauthoryear{Petty and Cacioppo}{Petty and
  Cacioppo}{1986}]%
        {petty1986elaboration}
\bibfield{author}{\bibinfo{person}{Richard~E Petty} {and}
  \bibinfo{person}{John~T Cacioppo}.} \bibinfo{year}{1986}\natexlab{}.
\newblock \showarticletitle{The elaboration likelihood model of persuasion}.
\newblock \bibinfo{journal}{{\em Advances in experimental social psychology\/}}
   \bibinfo{volume}{19} (\bibinfo{year}{1986}), \bibinfo{pages}{123--205}.
\newblock


\bibitem[\protect\citeauthoryear{Reno, Cialdini, and Kallgren}{Reno
  et~al\mbox{.}}{1993}]%
        {reno1993transsituational}
\bibfield{author}{\bibinfo{person}{Raymond~R Reno}, \bibinfo{person}{Robert~B
  Cialdini}, {and} \bibinfo{person}{Carl~A Kallgren}.}
  \bibinfo{year}{1993}\natexlab{}.
\newblock \showarticletitle{The transsituational influence of social norms}.
\newblock \bibinfo{journal}{{\em Journal of personality and social
  psychology\/}} \bibinfo{volume}{64}, \bibinfo{number}{1}
  (\bibinfo{year}{1993}), \bibinfo{pages}{104--112}.
\newblock


\bibitem[\protect\citeauthoryear{Schultz, Nolan, Cialdini, Goldstein, and
  Griskevicius}{Schultz et~al\mbox{.}}{2007}]%
        {schultz2007constructive}
\bibfield{author}{\bibinfo{person}{P~Wesley Schultz},
  \bibinfo{person}{Jessica~M Nolan}, \bibinfo{person}{Robert~B Cialdini},
  \bibinfo{person}{Noah~J Goldstein}, {and} \bibinfo{person}{Vladas
  Griskevicius}.} \bibinfo{year}{2007}\natexlab{}.
\newblock \showarticletitle{The constructive, destructive, and reconstructive
  power of social norms}.
\newblock \bibinfo{journal}{{\em Psychological science\/}}
  \bibinfo{volume}{18}, \bibinfo{number}{5} (\bibinfo{year}{2007}),
  \bibinfo{pages}{429--434}.
\newblock


\bibitem[\protect\citeauthoryear{Schwartz}{Schwartz}{1992}]%
        {schwartz1992universals}
\bibfield{author}{\bibinfo{person}{Shalom~H Schwartz}.}
  \bibinfo{year}{1992}\natexlab{}.
\newblock \showarticletitle{Universals in the content and structure of values:
  Theoretical advances and empirical tests in 20 countries}.
\newblock \bibinfo{journal}{{\em Advances in experimental social psychology\/}}
   \bibinfo{volume}{25} (\bibinfo{year}{1992}), \bibinfo{pages}{1--65}.
\newblock


\bibitem[\protect\citeauthoryear{Stray, F{\ae}gri, and Moe}{Stray
  et~al\mbox{.}}{2016}]%
        {stray2016exploring}
\bibfield{author}{\bibinfo{person}{Viktoria Stray}, \bibinfo{person}{Tor~Erlend
  F{\ae}gri}, {and} \bibinfo{person}{Nils~Brede Moe}.}
  \bibinfo{year}{2016}\natexlab{}.
\newblock \showarticletitle{Exploring norms in agile software teams}. In
  \bibinfo{booktitle}{{\em 17th International Conference on Product-Focused
  Software Process Improvement (PROFES)}}. Springer, \bibinfo{pages}{458--467}.
\newblock


\bibitem[\protect\citeauthoryear{Teh, Baniassad, Van~Rooy, and Boughton}{Teh
  et~al\mbox{.}}{2012}]%
        {teh}
\bibfield{author}{\bibinfo{person}{Alvin Teh}, \bibinfo{person}{Elisa
  Baniassad}, \bibinfo{person}{Dirk Van~Rooy}, {and} \bibinfo{person}{Clive
  Boughton}.} \bibinfo{year}{2012}\natexlab{}.
\newblock \showarticletitle{Social Psychology and Software Teams:
  {E}stablishing Task-Effective Group Norms}.
\newblock \bibinfo{journal}{{\em IEEE Software\/}} \bibinfo{volume}{29},
  \bibinfo{number}{4} (\bibinfo{year}{2012}), \bibinfo{pages}{53--58}.
\newblock


\bibitem[\protect\citeauthoryear{Terry and Hogg}{Terry and Hogg}{1996}]%
        {terry1996group}
\bibfield{author}{\bibinfo{person}{Deborah~J Terry} {and}
  \bibinfo{person}{Michael~A Hogg}.} \bibinfo{year}{1996}\natexlab{}.
\newblock \showarticletitle{Group norms and the attitude-behavior relationship:
  {A} role for group identification}.
\newblock \bibinfo{journal}{{\em Personality and Social Psychology Bulletin\/}}
  \bibinfo{volume}{22}, \bibinfo{number}{8} (\bibinfo{year}{1996}),
  \bibinfo{pages}{776--793}.
\newblock


\bibitem[\protect\citeauthoryear{Th{\o}gersen}{Th{\o}gersen}{2006}]%
        {thogersen2006norms}
\bibfield{author}{\bibinfo{person}{John Th{\o}gersen}.}
  \bibinfo{year}{2006}\natexlab{}.
\newblock \showarticletitle{Norms for environmentally responsible behaviour: An
  extended taxonomy}.
\newblock \bibinfo{journal}{{\em Journal of Environmental Psychology\/}}
  \bibinfo{volume}{26}, \bibinfo{number}{4} (\bibinfo{year}{2006}),
  \bibinfo{pages}{247--261}.
\newblock


\bibitem[\protect\citeauthoryear{Thurstone}{Thurstone}{1931}]%
        {thurstone1931measurement}
\bibfield{author}{\bibinfo{person}{Louis~Leon Thurstone}.}
  \bibinfo{year}{1931}\natexlab{}.
\newblock \showarticletitle{The measurement of social attitudes}.
\newblock \bibinfo{journal}{{\em The Journal of Abnormal and Social
  Psychology\/}} \bibinfo{volume}{26}, \bibinfo{number}{3}
  (\bibinfo{year}{1931}), \bibinfo{pages}{249}.
\newblock


\bibitem[\protect\citeauthoryear{Verplanken and Aarts}{Verplanken and
  Aarts}{1999}]%
        {verplanken1999habit}
\bibfield{author}{\bibinfo{person}{Bas Verplanken} {and} \bibinfo{person}{Henk
  Aarts}.} \bibinfo{year}{1999}\natexlab{}.
\newblock \showarticletitle{Habit, attitude, and planned behaviour: {I}s habit
  an empty construct or an interesting case of goal-directed automaticity?}
\newblock \bibinfo{journal}{{\em European review of social psychology\/}}
  \bibinfo{volume}{10}, \bibinfo{number}{1} (\bibinfo{year}{1999}),
  \bibinfo{pages}{101--134}.
\newblock


\bibitem[\protect\citeauthoryear{Wicker}{Wicker}{1969}]%
        {wicker1969attitudes}
\bibfield{author}{\bibinfo{person}{Allan~W Wicker}.}
  \bibinfo{year}{1969}\natexlab{}.
\newblock \showarticletitle{Attitudes versus actions: {T}he relationship of
  verbal and overt behavioral responses to attitude objects}.
\newblock \bibinfo{journal}{{\em Journal of Social issues\/}}
  \bibinfo{volume}{25}, \bibinfo{number}{4} (\bibinfo{year}{1969}),
  \bibinfo{pages}{41--78}.
\newblock


\end{thebibliography}

\end{document}